\documentclass[11pt,american,english,onecolumn]{IEEEtran}
\usepackage[T1]{fontenc}
\usepackage[latin9]{inputenc}
\usepackage{units}
\usepackage{amsmath}
\usepackage{graphicx}
\usepackage{setspace}
\usepackage{babel}

\begin{document}

\title{On Large Scale Distributed Compression and Dispersive Information
Routing for Networks}

\author{Kumar Viswanatha{*}, Sharadh Ramaswamy, Ankur Saxena, Emrah Akyol
and Kenneth Rose\\
\{kumar, rsharadh, ankur, eakyol, rose\}@ece.ucsb.edu \\
Department of Electrical and Computer Engineering \\
University of California at Santa Barbara, CA-93106.\thanks{The work was supported by the NSF under grants CCF-0728986, CCF - 1016861 and  CCF-1118075. The results in this paper were presented in part at the IEEE conference on acoustics speech and signal processing (ICASSP) 2010, Dallas, Texas, USA and at the European signal processing conference 2010, Aalborg, Denmark.} }
\maketitle
\begin{abstract}
This paper considers the problem of distributed source coding for
a large network. A major obstacle that poses an existential threat
to practical deployment of conventional approaches to distributed
coding is the exponential growth of the decoder complexity with the
number of sources and the encoding rates. This growth in complexity
renders many traditional approaches impractical even for moderately
sized networks. In this paper, we propose a new decoding paradigm
for large scale distributed compression wherein the decoder complexity
is explicitly controlled during the design. Central to our approach
is a module called the \textquotedblleft{}bit-subset selector\textquotedblright{}
whose role is to judiciously extract an appropriate subset of the
received bits for decoding per individual source. We propose a practical
design strategy, based on deterministic annealing (DA) for the joint
design of the system components, that enables direct optimization
of the decoder complexity-distortion trade-off, and thereby the desired
scalability. We also point out the direct connections between the
problem of large scale distributed compression and a related problem
in sensor networks, namely, dispersive information routing of correlated
sources. This allows us to extend the design principles proposed in
the context of large scale distributed compression to design efficient
routers for minimum cost communication of correlated sources across
a network. Experiments on both real and synthetic data-sets provide
evidence for substantial gains over conventional approaches. \end{abstract}

\section{Introduction}

The field of distributed source coding (DSC) has gained significant
importance in recent years, mainly due to its relevance to numerous
applications involving sensor networks. DSC originated in the seventies
with the seminal work of Slepian and Wolf \cite{Slepian_Wolf} where
they showed that, in the context of lossless coding, side-information
available only at the decoder can nevertheless be fully exploited
as if it were available to the encoder; in the sense that there is
no asymptotic performance loss. Later, Wyner and Ziv \cite{Wyner_Ziv}
derived a lossy coding extension. A number of theoretical publications
followed, primarily aimed at solving the general multi-terminal source
coding problem (see e.g. \cite{Han_Kobayashi}). It was not until
the late nineties, when first practical DSC schemes were designed
adopting principles from coding theory. Today the research in DSC
can be categorized into two broad camps. First approach builds on
methodologies from channel coding, wherein block encoding techniques
are used to exploit correlation \cite{DISCUS,Xiong,Turbo}. While
these techniques are efficient in achieving good rate-distortion performance,
they suffer from significant delays and high encoding complexities,
which make them unsuitable for many practical applications. The second
approach to DSC sprung directly from principles of source coding and
quantization \cite{Flynn-Gray,optimal_quant_DSC,NVQ,Ankur_DA,Yahampath_DPC,DPC}.
These techniques introduce low to zero delay into the system and typically
require minimal encoding complexity. Iterative algorithms for distributed
vector quantizer design have been proposed in \cite{optimal_quant_DSC,NVQ}.
A global optimization algorithm, based on deterministic annealing,
for efficient design of distributed vector quantizers was proposed
in \cite{Ankur_DA}. To optimize the fundamental trade-offs in the
(practically unavoidable) case of sources with memory have recently
appeared in \cite{Yahampath_DPC,DPC}. Source coding based approaches
will be most relevant to us here and will be discussed briefly in
Section \ref{sec:Conventional-Distributed-Source}. 

Distributed coding for a large number of sources is, in theory, a
trivial extension of the two source case, but the exponential codebook
size growth with the number of sources and encoding rates, makes it
nonviable in most practical applications. Just to illustrate, consider
$20$ sensors, transmitting information at $2$ bits per sensor. The
base station receives $40$ bits using which it estimates the observations
of all the sensors. This implies that the decoder has to maintain
a codebook of size $20\times2^{40}$, which requires about $175$
Terabytes of memory. In general, for $N$ sources transmitting at
$R$ bits, the total decoder codebook size would be $N2^{NR}$. Clearly,
the design of optimal low-storage distributed coders is a crucial
problem whose solution has existential ramifications to application
of DSC in real world sensor networks. Several researchers have addressed
this important issue in the past, e.g. \cite{factor_graphs,source_opt_clustering,Yahampath_BN}.
Most of these approaches are based on source grouping mechanisms where
the sources are first clustered based on their statistics and then
optimal DSC is designed independently within each cluster. However,
such approaches suffer from important drawbacks which will be explained
in detail in Section \ref{sec:Conventional-Distributed-Source}.

In this paper, inspired by our recent results in fusion coding and
selective retrieval of correlated sources \cite{Fusion_theory,Fusion_Coding,Pred_Fusion,Shared_des_fusion},
we propose a new decoding paradigm for large scale distributed coding,
where the design explicitly models and controls the decoding complexity.
Central to this approach is a new module called the bit-subset selector
that allows to judiciously select a subset of the received bits for
decoding each source. Specifically, to estimate each source, the bit-subset
selector selects only a subset of the received bits that provide the
most reliable estimate of the source. The decoder codebook size then
depends only the number of bits selected, which is explicitly controlled
during the design. Thus, a direct trade-off between decoder storage
complexity and reconstruction distortion is possible. We first present
a greedy iterative descent technique for the design of the encoders
along with the bit-subset selectors and the decoders and show significant
performance improvement over other state-of-the-art approaches. We
then present a deterministic annealing (DA) based global optimization
algorithm, due to the highly non-convex nature of the cost function,
which further enhances the performance over basic greedy iterative
descent technique. Experiments with both real and synthetic data sets
show that our approach reduces codebook complexity, by factors reaching
16X, over heuristic source grouping methods while achieving the same
distortion .

A different problem that is highly relevant to multi-hop sensor networks,
and that perhaps surprisingly reveals underlying conceptual similarity
with the problem of large scale distributed compression, is that of
routing correlated sources across a networks. We recently introduced
a new routing paradigm for sensor networks in \cite{ITW10,ISIT_2011_DIR}
called {}``dispersive information routing'' (DIR) and showed using
information theoretic principles that the new routing technique offers
significant (asymptotic) improvements in communication cost compared
to conventional routing techniques. In this paper, we point out the
close connection between the problem of large scale distributed compression
and dispersive information routing and thereby illustrate the general
applicability of the approaches herein. We then use similar principles
to design efficient low-delay dispersive information routers and demonstrate
using both, synthetic and real sensor network grids, that DIR offers
significant improvement in performance over conventional routing techniques.
We note that a preliminary version of our results appeared in \cite{our_ICASSP1}
and \cite{EUSIPCO}.

The contributions of this paper are summarized as follows:
\begin{enumerate}
\item We introduce a new decoding paradigm for large scale distributed coding
based on a new module at the decoder, called the bit-subset selector,
that judiciously selects a subset of the received bits to estimate
each source.
\item We propose a greedy iterative design strategy for the joint design
of the encoders, the bit-subset selectors and the decoders given a
training set of source samples and demonstrate using both real and
synthetic datasets that the proposed decoding paradigm provides significant
improvements in performance over conventional techniques. 
\item Motivated by the highly non-convex nature of the underlying cost function
during design, we propose a deterministic annealing based global optimization
approach for the joint design of the system parameters. The proposed
design algorithm provides additional gains in performance over the
greedy iterative algorithm.
\item We then relate the problem of large scale distributed compression
to that of multi-hop routing for sensor networks and based on the
underlying principles, we propose a new strategy for efficient design
of low-delay dispersive information routers. Simulation results on
datasets, collected from both real and synthetic sensor grids, show
significant gains over conventional routing techniques. 
\end{enumerate}
Much of the focus in the beginning of this paper will be on the problem
of large scale distributed compression. In Section \ref{sec:Conventional-Distributed-Source},
we begin by describing the canonical distributed source coder and
the challenges involved in scaling it to large networks. We then describe
the proposed decoding framework for large scale distributed compression
based on the bit-subset selector module. In Section \ref{sec:Algorithms-for-System},
we pose the design problem in a constrained optimization framework
that allows us to formulate a Lagrangian to quantify the codebook
complexity-distortion trade-off. We then propose an efficient algorithm
for the joint design of the system components based on deterministic
annealing and demonstrate significant gains over heuristic source
grouping methods. In Section \ref{sec:DIR}, we show that the proposed
methodology has wide applicability beyond the problem of large scale
distributed compression, where we extend the principles to design
a practical integrated framework for distributed compression and dispersive
information routing. We show that the bit-subset selector module and
the underlying design methodologies play a central role in designing
routers for minimum cost communication of correlated sources across
a network.

\section{Conventional Distributed Source Coder\label{sec:Conventional-Distributed-Source}}

\begin{figure}
\centering\includegraphics[scale=0.34]{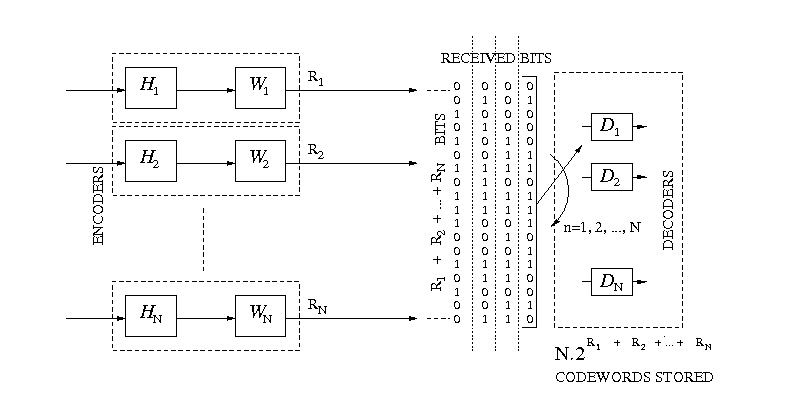}\caption{\label{fig:Conventional-DSC-Setup.}Conventional DSC Setup. Observe
that the decoder receives $\sum_{i}R_{i}$ bits and has to store a
unique reconstruction for every possible received index tuple.}

\end{figure}

Before describing the problem setup, we state some of the assumptions
made in this paper. First, as a simplifying assumption, we consider
only spatial correlations and ignore temporal correlations. However,
temporal correlations can be easily incorporated using techniques
similar to that in \cite{DPC}. Second, in this paper we assume that
the channels from the sensors to the sinks are noiseless/error-free.
The design of DSC in the presence of channel noise is an interesting
and challenging problem in its own right. Some preliminary research
in this directly appeared recently in \cite{IPSN}. In the first half
of the paper, we assume that there exists a separate link from every
source to the central receiver, i.e., information is not routed in
a multi-hop fashion. However, the method we propose is fairly general
and is applicable to the multi-hop setting. In the second half of
the paper, we focus on DSC design in conjunction with routing in multi-hop
sensor networks and show that the methodologies we develop play a
central role in efficient information dispersion across a network.
Throughout this paper, we make the practical assumption that while
the joint densities may not be known, there will be access to a training
sequence of source samples during the design. In practice this could
either be gathered off-line before the deployment of the network or
could be collected during an initial training phase. 

We begin with a description of the conventional DSC system with a
single sink. We refer to \cite{Ankur_DA} for a detailed description.
Consider $N$ correlated sources, $(X_{1},X_{2}\ldots,X_{N})$ transmitting
information at rate $R_{i}$, respectively, to the sink (fusion center).
The sink attempts to jointly reconstruct $(X_{1},X_{2}\ldots,X_{N})$
using bits received from all the sources. The setup is shown in Fig.
\ref{fig:Conventional-DSC-Setup.} as a block diagram. 

\begin{figure}
\centering\includegraphics[scale=0.34]{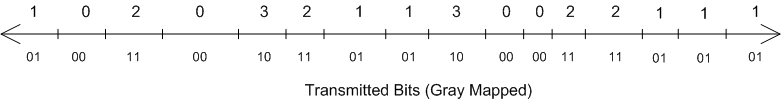}\caption{\label{fig:WZ_map}Example of a typical Wyner-Ziv Map}

\end{figure}

The encoding at each source consists of two stages. The first stage
is a simple high rate-quantizer which discretizes the input-space
into a finite number of regions $N_{i}$. Specifically, each high
rate quantizer is a mapping:\begin{equation}
\mathcal{H}_{i}:\mathcal{R}\rightarrow Q_{i}=\{1\ldots N_{i}\}\end{equation}
Note that these quantizers are made high rate so as to exclude them
from the joint encoder-decoder design. This is a practical engineering
necessity, (see e.g., \cite{Ankur_DA}), and the primary purpose of
the high rate quantizers is to discretize the source. The second stage,
called the \textquoteleft{}Wyner-Ziv map\textquoteright{} (WZ-map),
is a module that relabels the $N_{i}$ quantizer regions with a smaller
number, $2^{R_{i}}$, of labels. Mathematically, for each source $i$,
the WZ-map is the function,\begin{equation}
\mathcal{W}_{i}:Q_{i}\rightarrow\mathcal{I}_{i}=\{1\ldots2^{R_{i}}\}\end{equation}
and the encoding operation can be expressed as:\begin{equation}
\mathcal{E}_{i}(x_{i})=\mathcal{W}_{i}\left(\mathcal{H}_{i}\left(x_{i}\right)\right)\,\,\,\,\,\,\,\forall i\end{equation}
A typical example of a WZ-map is shown in Fig. \ref{fig:WZ_map}.
Observe that the WZ-maps make the encoding operation at each source
equivalent to that of an irregular quantizer wherein regions that
are far apart are mapped to the same transmission index. Although
this operation might seem counter intuitive at first, if designed
optimally, it is precisely these modules which assist in exploiting
correlations without inter-source communication. The decoder resolves
the ambiguity between the regions using the information received from
correlated sources. It is fairly well known (see \cite{Flynn-Gray})
that these WZ-maps, if properly designed, provide significant improvements
in the overall rate-distortion performance compared to that achievable
by regular quantizers operating at the same transmission rates. It
is important to note that the WZ-maps must be designed jointly using
the training sequence of observations, before the network begins its
operation. 

The decoder receives $\mathcal{E}_{i}(x_{i})=I_{i}\,\,\forall i$
as shown in the figure. We use $I=\{I_{1},I_{2},\ldots,I_{N}\}$ to
denote the received index tuple and $\mathcal{I}=\mathcal{I}_{1}\times\mathcal{I}_{2}\ldots\mathcal{I}_{N}$
to denote the set of all possible received symbols at the decoder.
The decoder $\mathcal{D}_{i}$ for each source at the sink is given
by a function:\begin{equation}
\mathcal{D}_{i}:\mathcal{I}\rightarrow\hat{\mathcal{X}}_{i}\in\mathcal{R}\end{equation}
Usually the decoder is assumed to be a look-up table, which has the
reconstruction values stored for each possible received index. For
optimal decoding, the look up table has a unique reconstruction stored
for each possible received index tuple and hence the total storage
at the decoder grows as $\mathcal{O}(N\times2^{R_{r}})=\mathcal{O}(N\times2^{\sum_{i=1}^{N}R_{i}})$%
\footnote{Note that even for a cascaded coding system, the look-up table at
each of the $N$ successive stages of the tree becomes exponentially
large.%
}, which is exponential in $N$. We refer to the total storage of the
look-up table as the decoder complexity. In most prior work, DSC was
performed for a few (typically 2 - 3) sources, with the implicit assumption
of design scalability with network size. But this exponential growth
in optimal decoder complexity with the number of sources and transmission
rates makes it infeasible to use the conventional setup in practical
settings even with moderately large number of sources. This is one
of the major obstacles that has deterred practical deployment of such
techniques in real world systems.

One natural solution proposed in the past to handle the exponential
growth in decoder complexity is to group the sources based on source
statistics \cite{source_opt_clustering} and to perform DSC within
each cluster. By restricting the number of sources within each group,
the decoder complexity is maintained at affordable limits. A major
difficulty with such an approach is to come up with good source grouping
mechanisms which are optimized for performing distributed compression.
While this problem is interesting in its own right, even if optimally
solved, such approaches do not exploit inter-cluster dependencies
and hence would lead to sub-optimal estimates. We will show in the
results section that the proposed bit-subset selector based decoding
paradigm provides significant improvements over such source grouping/clustering
techniques.

It is worthwhile to mention that an alternate approach, other than
the look-up table has been proposed in the literature to practically
implement the decoder \cite{factor_graphs,Yahampath_BN,Yahampath_MRF}.
In this approach, the decoder computes the reconstructions on the
fly by estimating the posterior probabilities for quantization index
$Q_{i}$ as $P(Q_{i}|I)$, when a particular index tuple is received.
Such an approach requires us to store the high rate quantization codewords
at the decoder, which grows only linearly in $N$. However, the computation
of the posterior probabilities $P(Q_{i}|I)$ requires an exponential
number of computations, let alone the exponential storage required
to store the joint pmf $P(Q_{1},Q_{2},\ldots,Q_{N})$. To limit the
computational complexity, prior work such as \cite{factor_graphs,Yahampath_BN,Yahampath_MRF}
have proposed clustering the sources and linking the clusters using
a limited complexity Bayesian network (or a factor graph), and thereby
using message passing algorithms to find $P(Q_{i}|I)$ with affordable
complexities. These approaches efficiently exploit inter-cluster dependencies
and hence provide significant improvement in distortion over simple
grouping techniques for fixed cluster sizes. However, a major shortcoming
of these approaches, which is often overlooked, is that the decoder
now also needs to store the Bayesian network/factor graph. Though
this storage grows linearly in $N$, it grows exponentially in the
`rate of the high rate quantizers'. Specifically, if $N_{i}=2^{R_{q}}\forall i$,
then the total storage at the decoder grows as $\mathcal{O}(N2^{MR_{q}})$,
where $M$ denotes the maximum number of parents for any source node
in the Bayesian network. We demonstrated recently in \cite{IPSN}
that this entails a considerable overhead in the total storage required
at the decoder and the gains in distortion are often overhauled by
the excess storage. We also demonstrated that it is indeed beneficial
to group more sources within a cluster instead of connecting the clusters
using a Bayesian network. Hence, in this paper, we compare the proposed
approach with just the source grouping technique, noting that, the
Bayesian network based approaches typically require significantly
higher storage to achieve good distortion performance. We refer to
\cite{IPSN} for a detailed analysis of the storage requirements for
the Bayesian network based techniques.

\section{Large Scale Distributed Compression - Proposed Setup\label{sec:LSDQ_Proposed}}

\begin{figure}
\centering\includegraphics[scale=0.46]{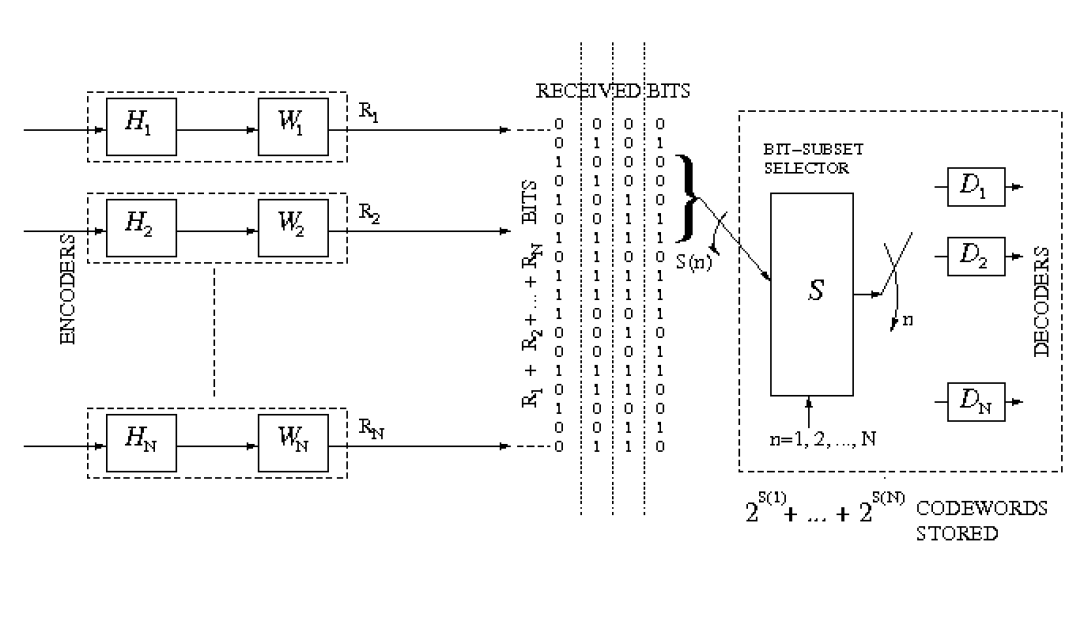}\caption{\label{fig:Proposed-setup-LSDQ}Proposed setup for large scale distributed
compression. The bit-subset selector judiciously selects a subset
of the received bits for decoding each source. Observe that the decoder
now need to only store $\sum_{i}2^{|\mathcal{S}(i)|}$ reconstruction
codewords.}

\end{figure}

We now describe our approach towards the design of large scale distributed
quantizers. Recall that the decoder receives $R_{r}=\sum_{i=1}^{n}R_{i}$
bits of information. For optimal decoding, the decoder needs to store
a unique reconstruction in the look up table for every possible received
combination of bits. This is infeasible as the decoder complexity
grows exponentially in $R_{r}$. Hence, we impose a structure on the
decoder wherein each source is decoded only based on a subset of the
received bits that provide the most reliable reconstruction for the
source. Essentially, we decompose the monolithic decoder, which was
a simple look-up table, by introducing an additional module called
the `bit-subset selector'. For each source, the bit-subset selector
determines the subset of the received bits that are used in reconstructing
the source. Note that the subset of bits used to estimate each source
could be different. Fig. \ref{fig:Proposed-setup-LSDQ} represents
our system as a block diagram.

Formally, the bit subset selector is the mapping :\begin{equation}
\mathcal{S}:\{1,...,N\}\rightarrow\mathcal{B}\in2^{\{1,\ldots,R_{r}\}}\label{eq:BS1}\end{equation}
where $\mathcal{B}$ is the power set (set of all possible subsets)
of the set $\{1,\ldots,R_{r}\}$. For source $i$, the bit-subset
selector retrieves the bits indicated by $\mathcal{S}(i)$ for decoding.
This implies that the decoder has to store a unique reconstruction
(codeword) for every bit tuple that could be selected. If we denote
the decoding rate by $R_{d_{i}}=|\mathcal{S}(i)|$ bits, the total
storage at the decoder is $\sum_{i=1}^{N}2^{R_{d_{i}}}$, where $|\cdot|$
denotes the set cardinality%
\footnote{Note that the decoder needs to also store the bit-subset selector
module along with the codebooks. However, it is easy to show that
the total storage required for the bit-subset selector is $(\sum_{i}R_{d_{i}})\log_{2}R_{r}$
bits and the overhead in storage is typically very small compared
to the memory required for the codebook storage.%
}. For different choices of $\mathcal{S}(i)$, reconstructions of source
$i$ at different distortion levels are possible and thus, reconstruction
quality can be traded against the required codebook size.

The decoder for each source is now modified to be the mapping: \begin{equation}
\mathcal{D}_{i}:\mathcal{I}\times\mathcal{B}\rightarrow\mathcal{R}\label{eq:Dec1}\end{equation}
and the reconstruction $\hat{X}_{i}=\mathcal{D}_{i}(I,\mathcal{S}(i))$,
where $\mathcal{D}_{i}(\cdot,\cdot)$ depends only on the bits in
$I$ at positions indicated by $\mathcal{S}(i)$. We now define the
decoder complexity as the average codebook size at the decoder given
by,\begin{equation}
C=\frac{1}{N}\sum_{i=1}^{N}2^{R_{d_{i}}}=\frac{1}{N}\sum_{i=1}^{N}2^{|\mathcal{S}(i)|}\label{eq:Complexity}\end{equation}
The average reconstruction distortion is measured as: \begin{equation}
D=E\left[\sum_{i=1}^{N}\gamma_{i}d_{i}(X_{i},\hat{X}_{i})\right]\label{eq:Dist1}\end{equation}
where $d_{i}:\mathcal{R}\times\mathcal{R}\rightarrow[0,\infty)$ $\forall i$
are well defined distortion measures and $0\leq\gamma_{i}\leq1:\sum_{i=1}^{N}\gamma_{i}=1$
are weights used to measure the relative importance of each of the
sources towards the total distortion. Hereafter, we will assume the
distortion to be the mean squared error (MSE) and assume equal importance
for all the sources, noting that the approach is applicable to more
general distortion measures and weights, i.e., \begin{equation}
D=\frac{1}{N}E\left[\sum_{i=1}^{N}(X_{i}-\hat{X}_{i})^{2}\right]\label{eq:Dist1_5}\end{equation}
In practice, we only have access to a training set and not the actual
source distributions. Hence, assuming ergodicity, we replace the $E[\cdot]$
operation by a simple empirical average over all the training samples.
If the training set is denoted by $\mathcal{T}$, we measure distortion
as:\begin{equation}
D=\frac{1}{N|\mathcal{T}|}\sum_{\mathbf{x}\in\mathcal{T}}\sum_{i=1}^{N}(x_{i}-\hat{x}_{i})^{2}\label{eq:Dist2}\end{equation}
where $\mathbf{x}=\{x_{1},\ldots,x_{N}\}$. We pose the design problem
in a constrained optimization framework wherein the objective is to
minimize the average distortion (averaged over the training set) subject
to a constraint on the total complexity at the decoder. The trade-off
between distortion and decoder complexity is controlled by a Lagrange
parameter $\lambda\geq0$ and by optimizing the weighted sum of the
two quantities. From (\ref{eq:Complexity}) and (\ref{eq:Dist2}),
the Lagrangian cost to be minimized is: \begin{eqnarray}
L & = & D+\lambda C\nonumber \\
 & = & \frac{1}{N|\mathcal{T}|}\sum_{\mathbf{x}\in\mathcal{T}}\sum_{i=1}^{N}(x_{i}-\mathcal{D}_{i}(I,S(i)))^{2}\nonumber \\
 &  & +\frac{\lambda}{N}\sum_{i=1}^{N}2^{|S(i)|}\label{eq:Lagrangian_formulation}\end{eqnarray}
Our objective is to find the optimal encoders (WZ-maps), the bit-subset
selectors and the reconstruction codebooks that minimize $L$ for
any given value of $\lambda$, i.e., \begin{equation}
\min_{\mathcal{E}_{i},\mathcal{D}_{i},\mathcal{S}}\,\,\,\, L=D+\lambda C\label{eq:Lag1}\end{equation}

\section{Algorithms for System Design\label{sec:Algorithms-for-System}}

A natural approach to design such systems is to first derive necessary
conditions for optimality of each of the modules and then to iteratively
optimize the different modules following a greedy gradient-descent
approach. However (\ref{eq:Lag1}) is a highly non-convex function
of the system parameters which makes the greedy approach very likely
to get trapped in poor local minima (even when optimized over multiple
random initializations), thereby leading to sub-optimal designs. Finding
a good initialization for such greedy iterative descent algorithms,
even for problems much simpler in nature than the one at hand, is
known to be a difficult task. Hence, in this section, we derive a
global optimization technique based on deterministic annealing (DA)
\cite{DA}, which has proven to be highly effective in solving related
problems in compression and classification. Nevertheless, we begin
by describing a greedy-descent algorithm for illustration and ease
of understanding. We note that the high-rate quantizers are designed
independently (using a standard quantizer design algorithm) and are
excluded from the joint design of the remaining system parameters.

\subsection{Greedy-Descent Algorithm}

\subsubsection{Necessary Conditions for Optimality}

We first derive the necessary conditions for optimality of all modules
in the proposed approach to distributed coding. 

\textbf{1) Optimal Encoders:} Let $\mathcal{T}_{i,j}=\{\textbf{x}\in\mathcal{T}:\mathcal{H}_{i}(x_{i})=j\}$.
Then, from (\ref{eq:Lag1}), the optimal WZ-map, given fixed bit-subset
selector and decoder codebooks is: \begin{eqnarray}
\mathcal{W}_{i}(j)=\arg\min_{k\in\mathcal{I}_{i}}\sum_{\mathbf{x}\in\mathcal{T}_{i,j}}\sum_{l=1}^{N}(x_{l}-\mathcal{D}_{l}(I_{i,k},S(l)))^{2}\label{eq:enc_update1-1}\end{eqnarray}
 where\[
I_{i,k}=[\mathcal{E}_{1}(x_{1}),\ldots,\mathcal{E}_{i-1}(x_{i-1}),k,\mathcal{E}_{i+1}(x_{i+1}),\ldots,\mathcal{E}_{N}(x_{N})]^{T}\]

\textbf{2) Optimal Bit-Subset Selector: }For fixed encoders and decoder
codebooks, the optimal subset of bits to be used to estimate each
source is given by:\begin{equation}
\mathcal{S}(i)=\arg\min_{e\in\mathcal{B}}\frac{1}{N|\mathcal{T}|}\sum_{\mathbf{x}\in\mathcal{T}}(x_{i}-\mathcal{D}_{i}(I,e))^{2}+\frac{\lambda}{N}\times2^{|e|}\label{eq:bs_update-1}\end{equation}

\textbf{3) Optimal Decoders :} If $I=[\mathcal{E}_{1}(x_{1}),\mathcal{E}_{2}(x_{2})\ldots\mathcal{E}_{N}(x_{N})]^{T}$
represents the bits received from all the sources, and if $e$ represents
the positions of bits selected by the bit-subset selector, then we
use $I_{e}$ to denote the index obtained by extracting the bits in
$I$ at the positions indicated by $e$. Differentiating the expression
for $L$, (\ref{eq:Lag1}), with respect to the reconstruction values
and equating it to zero gives the optimal expression for the decoder
to be:\begin{equation}
\hat{x}_{i}(I_{e})=\mathcal{D}_{i}(I,e)=\frac{1}{|\mathcal{F}|}\sum_{\mathbf{x}\in\mathcal{F}}x_{i}\label{eq:dec_update-1}\end{equation}
where $\mathbf{x}\in\mathcal{F}$ if $(\{\mathcal{E}_{1}(x_{1}),\mathcal{E}_{2}(x_{2})\ldots\mathcal{E}_{N}(x_{N})\})_{e}=I_{e}$.

Given the necessary conditions for optimality, a natural design rule
is to iteratively optimize the different modules. The algorithm begins
with random initializations of all the system parameters. Then, all
the parameters are iteratively optimized using (\ref{eq:enc_update1-1}),
(\ref{eq:bs_update-1}) and (\ref{eq:dec_update-1}) until convergence.
When each module is optimized, it leads to a reduction in the Lagrangian
cost. Since there are only a finite number of partitions of the training
set, convergence to a locally optimal solution is guaranteed. By varying
$\lambda$, the trade-off between decoder complexity and distortion
is controlled and an operational complexity-distortion curve is obtained.
The system is optimized over multiple random initializations to avoid
poor local minima. However, as we will show in the following section,
an algorithm based on DA performs significantly better than such a
greedy-descent technique, even with multiple random initializations.

\subsubsection{Low Complexity Search}

Before we describe the DA based algorithm for system design, there
is a caveat that needs to be addressed. First, let us consider the
design complexity of WZ maps. The WZ-maps are updated using (\ref{eq:enc_update1-1})
which entails a complexity of $\mathcal{O}(N|\mathcal{T}|)$ per source.
Hence, the the overall complexity of updating WZ-maps in each iteration
grows as $\mathcal{O}(N^{2}|\mathcal{T}|)$, which is quadratic in
the number of sources and hence is assumed affordable%
\footnote{Note that, before iterative optimization, the design of the high rate
quantizers for all source is performed using the Lloyd-Max scheme,
which is a relatively low complexity step.%
}. However, the design of the bit-subset selector involves considerably
higher complexity. In order to find the best bit-subset selector,
every possible configuration of the bit-subset selector must be considered
for decoding each source. This will necessitate $\mathcal{O}(N2^{R_{r}})$
calculations and entails a storage of $N\times\sum_{k=1}^{R_{r}}({R_{r}\atop k})2^{k}=N(3^{R_{r}}-1)$
codewords during design. Recall that $R_{r}$ grows linearly in $N$,
thus making the net complexity cost of updating the bit-subset selector,
exponential in $N$. Hence, for the design algorithm described above,
the number of computations and codewords to be maintained grows exponentially
in $N,$ which makes it infeasible to implement in practice. Thus
we resort to a heuristic approach wherein, instead of finding the
best among all possible bit-subset selectors in each iteration of
the optimization, we choose an incrementally better solution. Specifically,
the search is restricted to only those bit-subsets that differ from
the current bit-subset selector setting in exactly one position, i.e.,
we restrict the search to $R_{r}$ bit subsets at a Hamming distance
of $1$ from the current solution. It is easy to verify that the design
complexity of such a heuristic technique is $\mathcal{O}(N^{2}R_{r})$,
which is only cubic in $N$. An added advantage is that this approach
requires far smaller storage during design as during each iteration,
only $R_{r}$ codebooks need to be maintained. As we will see in Section
\ref{sec:Results_LSDQ}, such a low-complexity design technique performs
very close to the full-complexity search for small networks indicating
that the loss in optimality due to such a heuristic approach is minimal.

\subsection{Deterministic Annealing Based Design}

In this section, motivated by the highly non-convex nature of the
cost function in (\ref{eq:Lagrangian_formulation}), we derive a deterministic
annealing (DA) based algorithm for the system design. The approach
derives its principles from \cite{DA} and builds upon the DA derivation
proposed in \cite{Ankur_DA} to encompass the design of the bit-subset
selectors jointly with the design of the WZ-maps and the reconstruction
codebooks. 

A formal derivation of the DA algorithm is based on principles borrowed
from information theory and statistical physics. Here, during the
design stage, we cast the problem in a probabilistic framework, where
the standard deterministic WZ-maps are replaced by a probabilistic
mapping, i.e., each of the $N_{i}$ regions associated with the high
rate quantizers are now assigned to each of the $2^{R_{i}}$ transmission
indices in probability. The expected cost is then minimized subject
to an entropy constraint that controls the \textquotedblleft{}randomness\textquotedblright{}
of the solution. By gradually relaxing the entropy constraint we obtain
an annealing process that seeks the minimum cost solution. It is important
to note that the WZ-mappings are made soft only during the design
stage. Of course, our final objective is to design hard WZ-mappings
which minimize the average Lagrangian cost for a fixed $\lambda$.
More detailed derivation and the principle underlying DA can be found
in \cite{DA}.

In the proposed design approach, we seek to optimize the WZ-maps and
the reconstruction codebooks for a fixed bit-subset selector using
DA. The bit-subset selectors are then updated to an incrementally
better solution using the low complexity update step described in
the previous section. This process is repeated until convergence.
Hence, during each iteration of the annealing process, the bit-subset
selector, and hence the decoder codebook complexity, is fixed. Thus
the Lagrangian cost in (\ref{eq:Lag1}) is determined only by the
average distortion, while the codebook complexity term can be temporarily
neglected from the cost function. Specifically, during the design,
we map the quantization region $q_{i}\in Q_{i}=\{1,\ldots,N_{i}\}$
to the transmission index $k_{i}\in\mathcal{I}_{i}=\{1,\ldots,2^{R_{i}}\}$
with probability $P_{i}(k_{i}|q_{i})$ $\forall i\in\{1,2,\ldots,N\}$.
This implies that each of the training set samples are assigned to
the transmission indices in probability determined by: \begin{equation}
P(k_{1},k_{2},\ldots,k_{N}|x_{1},x_{2},\ldots,x_{N})=\prod_{i=1}^{N}P_{i}(k_{i}|x_{i})\label{eq:DA1}\end{equation}
$\forall k_{i}\in\mathcal{I}_{i}$, $\forall(x_{1},x_{2},\ldots,x_{N})\in\mathcal{T}$,
where\begin{equation}
P_{i}(k_{i}|x_{i})=P_{i}(k_{i}|\mathcal{H}_{i}(x_{i}))\label{eq:DA2}\end{equation}
We note that the independence assumption in (\ref{eq:DA1}) is required
to ensure that the WZ-maps obtained at the end of the annealing process
operate independently at the respective encoders. The average distortion
is now measured as:

\begin{eqnarray}
D_{avg} & = & \frac{1}{N|\mathcal{T}|}\sum_{\mathbf{x}\in\mathcal{T}}\sum_{k_{1},k_{2}\ldots k_{N}}\Bigl\{ p(k_{1},k_{2},\ldots k_{N}|\mathbf{x})\nonumber \\
 &  & \times\sum_{i=1}^{N}(x_{i}-\mathcal{D}_{i}(k_{1},k_{2},\ldots k_{N},\mathcal{S}(i))^{2}\Bigr\}\label{eq:DA3}\end{eqnarray}
We seek to minimize the above average distortion subject to a constraint
on the average randomness in the system. The system randomness is
measured using the Shannon's entropy function given by:\begin{equation}
H=-\frac{1}{N|\mathcal{T}|}\sum_{\mathbf{x}\in\mathcal{T}}\sum_{i=1}^{N}\sum_{k_{i}\in\mathcal{I}_{i}}P(k_{i}|x_{i})\log\left(P(k_{i}|x_{i})\right)\label{eq:DA4}\end{equation}

The DA algorithm minimizes $D_{avg}$ in (\ref{eq:DA3}) (for a fixed
bit-subset selector), with a constraint on the entropy of the system,
(\ref{eq:DA4}), where the level of randomness is controlled by a
Lagrange parameter (usually called the temperature of the system due
to its roots in statistical physics), $T$ as:\begin{equation}
J=D_{avg}-TH\label{eq:DA5}\end{equation}
Initially, when $T$ is set very high, out objective is to maximize
the entropy of the system and hence the quantization regions are mapped
to all the the transmission indices with equal probability. Then during
each iteration of the annealing process, the temperature is gradually
lowered maintaining $J$ at its minimum. For example, during each
iteration, $T$ is updated as $T\rightarrow\alpha T$, where $\alpha=0.9$.
At each temperature, the probabilities $P_{i}(k_{i}|q_{i})$ and the
reconstruction codebooks are perturbed and then iteratively updated
until the system reaches equilibrium. Towards deriving the update
equations for the probabilities, note that (\ref{eq:DA5}) is convex
in $P_{i}(k_{i}|q_{i})$ and hence $\forall k_{i}\in\{1,\ldots,2^{R_{i}}\},$
$q_{i}\in\{1,\ldots,N_{i}\}$, the optimum expression for $P(k_{i}|q_{i})$
is given by the following Gibbs distribution (refer to \cite{Ankur_DA,DA}
for the details):\begin{equation}
P(k_{i}|q_{i})=\frac{e^{-\frac{D_{q_{i},k_{i}}}{T}}}{\sum_{k_{i}\in\mathcal{I}_{i}}e^{-\frac{D_{q_{i},k_{i}}}{T}}}\label{eq:Update_DA_1}\end{equation}
where \begin{eqnarray}
D_{q_{i},k_{i}} & = & E\left[D\Bigl|\mathcal{H}_{i}(x_{i})=q_{i},\, p(k_{i}|q_{i})=1\right]\label{eq:DA6}\\
 & = & \sum_{\mathbf{x}\in\mathcal{T}:\mathcal{H}_{i}(x_{i})=q_{i}}\sum_{k_{1},\ldots,k_{N}}\Bigl\{\prod_{j=1,j\neq i}^{N}P_{j}(k_{j}|x_{j})\nonumber \\
 &  & \times\sum_{j=1}^{N}(x_{j}-\mathcal{D}_{j}(k_{1},k_{2},\ldots k_{N},\mathcal{S}(j))^{2}\Bigr\}\end{eqnarray}
i.e., $D_{q_{i},k_{i}}$ is the average distortion if the quantization
region $q{}_{i}$ is mapped to the transmission index $k_{i}$ deterministically.
For fixed $P(k_{i}|q_{i})$, the reconstruction codewords are updated
using the following equation:\begin{equation}
\hat{x}_{i}(I_{e})=\frac{1}{|\mathcal{T}|}\sum_{\mathbf{x}\in\mathcal{T}}P(I_{e}|\mathbf{x})x_{i}\label{eq:DA7}\end{equation}
where $I_{e}$ denotes the subset of the received bits at positions
denoted by $e$. 

Hence at each temperature, the reconstruction codebooks and the probabilities
are updated iteratively using (\ref{eq:Update_DA_1}) and (\ref{eq:DA7})
till an equilibrium is reached. Finally as $T\rightarrow0$, $P(k_{i}|q_{i})$
in (\ref{eq:Update_DA_1}) converge to hard mappings and we obtain
the optimum WZ-maps and the codebooks that minimize $L$ for the given
bit-subset selector. 

\textbf{Note on Complexity:} Observe that, if we use (\ref{eq:DA3})
to compute the average distortion, the summation over all $k_{1},k_{2}\ldots k_{N}$
requires an exponential number of computations to be performed, i.e.,
the design complexity of DA grows as $\mathcal{O}(N2^{R_{r}})$. However,
a simple trick that exploits the fact that the bit-subset selector
is deterministic while the randomness is all in the WZ maps, allows
us to solve this problem. Specifically, the distortion in (\ref{eq:DA3})
can be expressed alternatively as:\begin{equation}
D=\frac{1}{N|\mathcal{T}|}\sum_{\mathbf{x}\in\mathcal{T}}\sum_{i=1}^{N}\sum_{s_{i}\in\mathcal{I}(\mathcal{S}(i))}P(s_{i}|\mathbf{x})\left(\hat{x}_{i}(s_{i})-x_{i}\right)^{2}\label{eq:DA8}\end{equation}
where $\mathcal{I}(\mathcal{S}(i))$ denotes the set of all index
tuples that can be selected by the bit-subset selector to decode source
$i$ and hence the cardinality of $\mathcal{I}(\mathcal{S}(i))$ is
$2^{|\mathcal{S}(i)|}$. Observe that the above operation requires
far fewer compuations, requiring a complexity of $\mathcal{O}(|\mathcal{T}|\sum_{i=1}^{N}2^{|\mathcal{S}(i)|})$,
which is only linear in $N$. We further note that computation of
$D_{q_{i},k_{i}}$ is effectively an operation that involves finding
distortions and hence the complexity of updating the probabilities
in (\ref{eq:Update_DA_1}) grows quadratic in $N$. Further, the codebook
update operation in (\ref{eq:DA7}) is a low complexity operation
requiring computations only of the order of $N$. Thus the overall
complexity of the DA based algorithm grows of the same order as that
for the greedy iterative descent algorithm discussed earlier. We further
note that the design is a one-time process and needs to be performed
only once before operation. The computational complexity during run-time
is negligible and is independent of the algorithm used for the design.

\section{Results for Large Scale Distributed Compression\label{sec:Results_LSDQ}}

We tested our proposed algorithm extensively on both synthetic and
real sensor network data and obtained the operational complexity (C)
vs. the distortion (D) (C-D curve) for each of the data sets%
\footnote{We note that the well known `time sharing\textquoteright{} argument
is not applicable to the complexity-distortion trade-off curve. This
is because, if we time share between two operating points, the decoder
needs to maintain large enough memory required by the operating point
with higher complexity. Nevertheless, we continue to connect all the
operating points for clarity of presentation.%
}. In all our simulations we considered a transmission rate $R_{i}=2$bits
per source and $\gamma_{i}=\frac{1}{N}$. The high rate quantizer
partitioned the input space into $32$ regions. To be fair, we used
the same high rate quantizers for the proposed technique and the competitors.
For all greedy-descent based design methods, we report the best performance
over several random initializations (limited to 25). The maximum average
complexity allowed for the decoder codebook was $1024$ ($10$ bits)%
\footnote{Note that the smallest complexity we consider is $4$, as we force
the bit-subset selector to select at least the $2$ bits sent by the
corresponding encoders.%
}.

As a competitor to our approach, we grouped the sources heuristically
based on their correlations ensuring that highly correlated sources
are grouped together and then independently applied conventional distributed
coding within each group. We varied the number of sources in each
group to obtain the distortion at different complexities. We considered
three data-sets for our analysis:

1) \textbf{5 Synthetic Gaussian sources} : We first considered 5 uniformly
spaced synthetic Gaussian sources, $\mathcal{N}(0,1)$, with correlation
exponentially decaying with the distance. Specifically, the correlation
coefficient between sources $X_{i}$ and $X_{j}$, is $\rho_{ij}=\rho^{|i-j|}$.
In our simulations, we assumed $\rho=0.95$ and a training set of
length $|\mathcal{T}|=200,000$. The results reported are on a test
set, also of length $200,000$.

2) \textbf{10 Synthetic Gaussian sources} : As a second data set,
we considered 10 uniformly spaced synthetic Gaussian sources, $\mathcal{N}(0,1)$,
again with an exponentially decaying correlation coefficient, with
$\rho=0.95$. We again chose $|\mathcal{T}|=200,000$ for both the
training and the test sets.

3) \textbf{50 Real world sensors data}: The real sensor network dataset
we used was collected by the Intel Berkeley Research Lab, CA%
\footnote{The dataset is available for download at : http://db.csail.mit.edu/labdata/labdata.html%
}. Data were collected from 54 sensors deployed in the Intel Berkeley
Research Lab between February 28 and April 5, 2004. The dataset had
recoding of temperature and luminescence from 25 sensors that collected
the highest number of samples, which is equivalent to 50 sources.
Each sensor recorded reading once every 31 s. Times when subset of
these sensors failed to record data were dropped from the analysis.
The data were normalized to zero mean and unit variance. Half the
dataset was used for training and the remaining for testing.

\selectlanguage{american}%
\begin{figure}
\selectlanguage{english}%
\centering\includegraphics[scale=0.27]{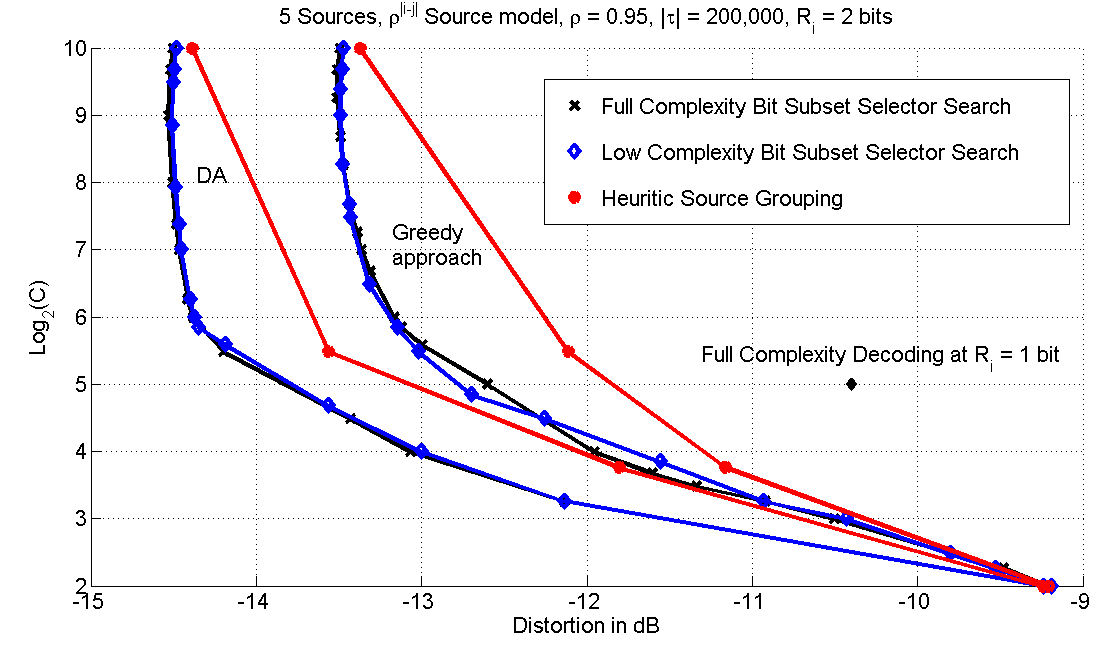}\foreignlanguage{american}{\caption{\selectlanguage{english}%
Complexity versus distortion trade-off for 5 Synthetic Gaussian sources
\label{fig:5_LSDQ}\selectlanguage{american}
}
}\selectlanguage{american}

\end{figure}

\begin{figure}
\selectlanguage{english}%
\centering\includegraphics[scale=0.27]{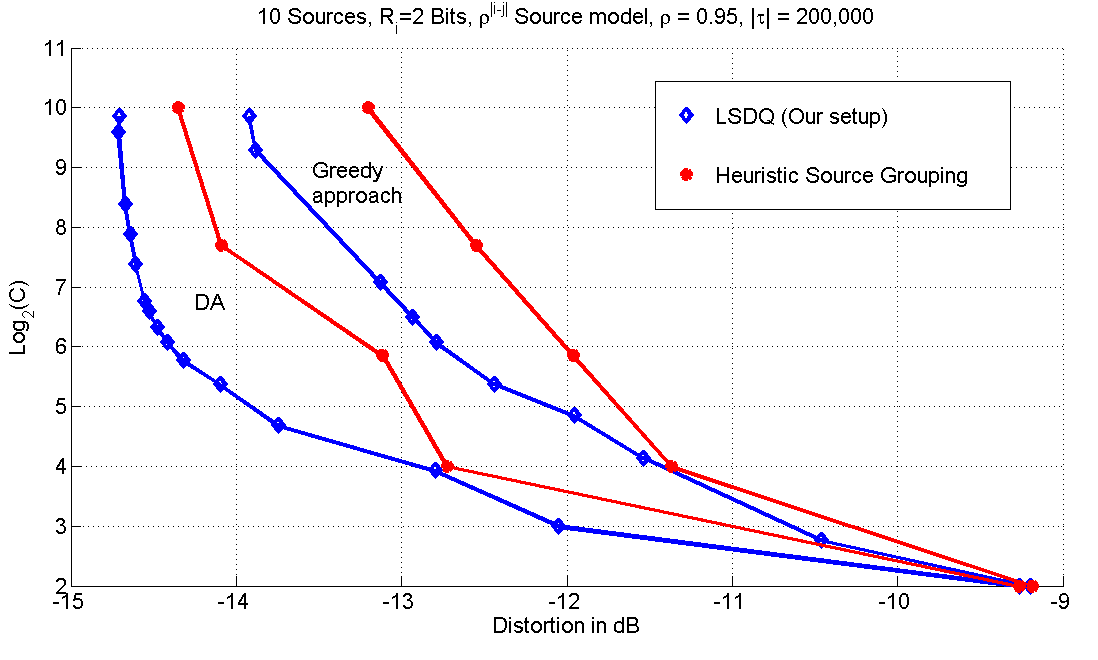}\foreignlanguage{american}{\caption{\selectlanguage{english}%
Complexity versus distortion trade-off for 10 Synthetic Gaussian sources\foreignlanguage{american}{\label{fig:10_LSDQ}}\selectlanguage{american}
}
}\selectlanguage{american}

\end{figure}

\begin{figure}
\selectlanguage{english}%
\centering\includegraphics[scale=0.27]{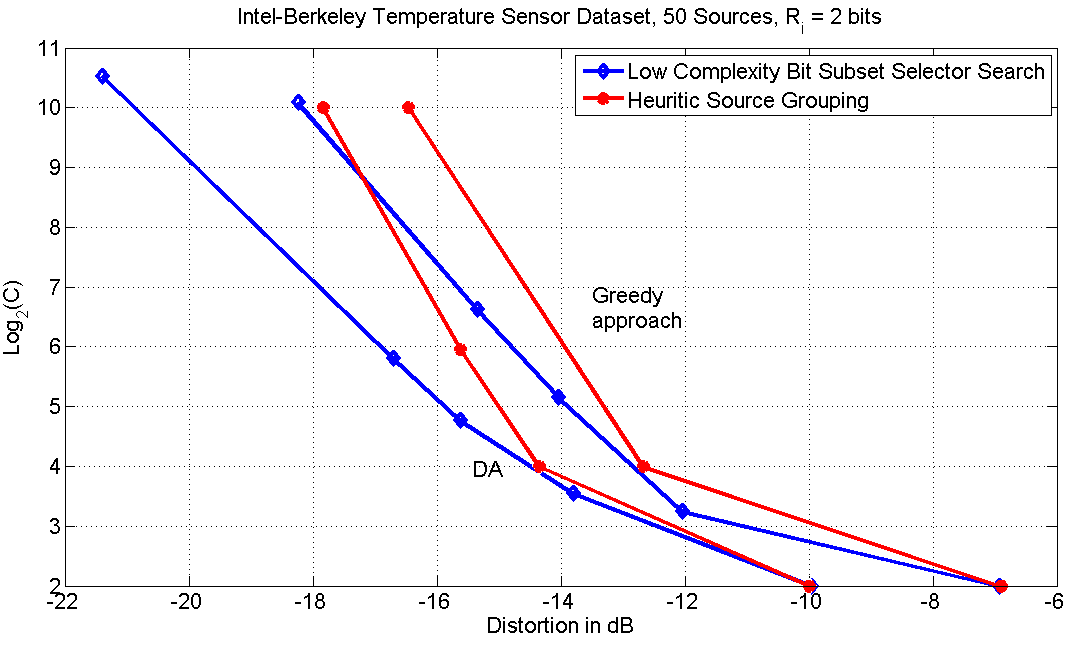}\foreignlanguage{american}{\caption{\selectlanguage{english}%
Complexity versus distortion trade-off for Intel-Berkeley temperature
sensor dataset - 50 sources\foreignlanguage{american}{\label{fig:Berk_LSDQ}}\selectlanguage{american}
}
}\selectlanguage{american}

\end{figure}

\selectlanguage{english}%
The complexity-distortion trade-off curve obtained by optimizing the
system using both greedy-iterative decent and using DA are shown in
figures \ref{fig:5_LSDQ}, \ref{fig:10_LSDQ} and \ref{fig:Berk_LSDQ},
for all the three datasets, respectively. To be on fair grounds, we
also report the results obtained by optimizing the source grouping
approach using deterministic annealing. Observe that, for the 5 source
synthetic dataset, it is possible to design the full complexity decoder
which operates at a decoder complexity of $2^{10}$. For this dataset,
our approach (optimized using greedy-iterative decent) outperforms
the heuristic grouping scheme by about 1 dB at a complexity of $2^{7}$.
Further gains of close to $3$ dB in distortion are possible using
the proposed design based on DA. We also report the performance of
the conventional distributed source coder transmitting at a lower
rate of $R_{i}=1$ bit per source. The proposed approach gains close
to $2.5$ dB in distortion at the same decoding complexity compared
to the conventional DSC transmitting at a lower rate. This demonstrates
that it is indeed beneficial to transmit more bits and allow the decoder
to select a subset of the received bits for decoding, rather than
to transmit fewer bits. 

Similar gains are seen for the other two datasets. For the 10 source
synthetic Gaussian dataset, we observe close to $1$ dB gain in distortion
at an average complexity of $2^{6}$ and a further improvement of
$2$ dB using deterministic annealing based design of the system parameters.
For the real world temperature sensor dataset, we see gains of close
to $2$ dB in distortion at a complexity of $2^{10}$ and a further
improvement of $3$ dB using deterministic annealing. Alternatively,
we see a $16$x reduction in decoder complexity at a distortion of
$-16$ dB.

\section{Distributed Source Coding and Dispersive Information Routing \label{sec:DIR}}

In this section, we consider a seemingly different application involving
minimum cost communication of correlated sources across a network
and show the broad applicability of the bit-subset selector module
and the design methodologies we proposed in the context of large scale
distributed compression. Recall the functionality of the bit-subset
selector module we described in Section \ref{fig:Proposed-setup-LSDQ}.
Its primary purpose was to select a subset of the transmitted bits
that provide the most reliable estimate for each source. By controlling
the number of bits selected, the decoder complexity was traded for
the reconstruction distortion. This intuition finds applicability
is several scenarios beyond the problem of large scale distributed
compression. In fact, the motivation for using the bit-subset selector
module was originally derived from its applications in databases for
fusion storage and selective retrieval of correlated sources \cite{Fusion_Coding}.
Here, we focus on a seemingly different application called dispersive
information routing (DIR) which involves communicating (routing) correlated
sources across a network. Exactly the same underlying principles will
be applicable here for efficient joint design of distributed source
coders and dispersive information routers. It is important to remember
that, while we demonstrate the applicability in the context of routing
in sensor networks, the underlying principles can be applied to a
rich class of problems involving `fragmentation of information'. 

Compression in multi-hop networks has gained significant importance
in recent years, mainly due to its relevance in sensor network applications.
Review paper \cite{Luo} describes the important joint compression-routing
schemes that have been developed so far. Encoding correlated sources
in a network with multiple sources and sinks has conventionally been
looked at from two different directions. The first approach performs
compression at intermediate nodes \cite{pattam_08}, where all the
information is available, without appealing to DSC principles. However,
such approaches tend to be wasteful at all but the last hops of the
communication path. The second approach uses distributed source coding
to exploit correlation at the source nodes followed by simple routing
at intermediate nodes. Well designed DSC could provide considerable
performance improvement and/or complexity/energy savings. Various
aspects of DSC for routing have been considered in a number of publications,
and notably in \cite{Networked_Slepian_Wolf}, where the authors consider
joint optimization of Slepian - Wolf coding and conventional routing.

Optimal routing schemes, designed for independent sources (conventional
routing), have been studied extensively \cite{Cormen}, primarily
due to the growth of the Internet. It may be tempting to assume that
an optimal distributed source code, which tries to eliminate the dependencies
between sources, followed by a conventional routing mechanism, would
achieve optimality in communication cost for transmitting correlated
sources across a network (see e.g. \cite{Networked_Slepian_Wolf,distance_entropy}).
However, we showed recently in \cite{ISIT10} that for a network with
multiple sources and sinks, DSC followed by conventional routing suffers
from inherent and significant drawbacks. We then introduced a new
paradigm called \textquotedblleft{}dispersive information routing\textquotedblright{}
(DIR) in \cite{ITW10,ISIT_2011_DIR}, which provides significant asymptotic
gains compared to conventional routing for communicating correlated
sources. In this new routing paradigm, every intermediate node is
allowed to split a packet and forward only a subset of received bits
on each of the forward paths. In this paper, deriving principles from
the problem of large scale distributed compression, we propose a practical
integrated framework for distributed compression and dispersive information
routing and show that the bit-subset selector module and the design
methodologies we proposed in the context of large scale distributed
compression, play a central role in designing such routers. We first
show using a simple network example the sub-optimality of conventional
routing methods and motivate the new paradigm. We then formulate the
problem of joint design of DSC and DIR that allows us to use similar
algorithms as in Section \ref{sec:Algorithms-for-System}. We then
apply the design to a sensor grid with multiple sources and sinks
and demonstrate the potential gains of the proposed methodology. We
note that unlike network coding \cite{DSC_NC,NC_Cost}, DIR can be
realized using conventional routers without recourse to expensive
coders at intermediate nodes, making it more suitable for low powered
sensor networks. We begin with the problem setup followed by a brief
information theoretic motivation for the new routing paradigm. We
refer to \cite{ISIT_2011_DIR} for a detailed analysis on the asymptotic
gains achievable using DIR.

\subsection{Problem Setup\label{sub:DIR_Problem-Setup}}

Let a network be represented by an undirected graph $G=(V,E)$. Each
edge $e\in E$ is a network link whose communication cost depends
on the edge weight $w_{e}$. The nodes $V$ consist of $N$ source
nodes, $M$ sinks, and $|V|-N-M$ intermediate nodes. Source node
$i$ observes $X_{i}$ where $(X_{1},X_{2},\ldots,X_{N})$ are correlated
random variables. The sinks are denoted $S_{1},S_{2},\ldots,S_{M}$.
Each sink requests information from a subset of sources. Let the subset
of nodes requested by sink $S_{j}$ be $V_{j}\subseteq V$. Conversely,
source $i$ has to be reconstructed at a subset of sinks denoted $S_{i}\subseteq\{S_{1},S_{2},\ldots,S_{M}\}$.
Define traffic matrix (or \textquotedblleft{}request\textquotedblright{}
matrix) $T$, for network graph $G$ as the $N\times M$ binary matrix
that specifies which sources must be reconstructed at each sink:\begin{equation}
T_{ij}=\begin{cases}
1 & \mbox{if }i\in V_{j}\\
0 & \mbox{else }\end{cases}\label{eq:T_defn}\end{equation}
Without loss of generality we assume that every source is requested
by at least 1 sink. 

The cost of communication through a link is a function of the bit
rate flowing through it and the edge weight, which we will assume
for simplicity to be a simple product $f(R,w_{e})=Rw_{e}$, noting
that the approach is directly extendible to more complex cost functions.
The objective is to design encoders, routers and decoders to minimize
the overall network cost (calculated given the set of link rates and
edge weights) at a prescribed level of average distortion.

We denote by $E_{B}^{i}$, the set of all paths from source $i$ to
the subset of sinks $B\subseteq\{S_{1},S_{2},\ldots,S_{M}\}$. The
optimum route from the source to these sinks is determined by a spanning
tree optimization (minimum Steiner tree) \cite{Cormen}. More specifically,
for each source node $i$, the optimum route is obtained by minimizing
the cost over all trees rooted at node $i$ which span all sinks $S_{j}\in B$.
The minimum cost of transmitting a single bit from source $i$ to
the subset of sinks $B$, denoted by $d_{i}^{*}(B)$, is given by:\[
d_{i}^{*}(B)=\min_{P\in E_{B}^{i}}\sum_{e\in P}w_{e}\]
These costs will play an important role in the proposed joint DSC-DIR
design in Section \ref{sub:Integrated-DSC-and}. However, before we
describe the proposed framework, we begin with an information theoretic
motivation for the new routing paradigm.

\subsection{Information Theoretic Motivation\label{sub:Information-Theoretic-Motivation}}

\begin{figure}
\centering\includegraphics[scale=0.3]{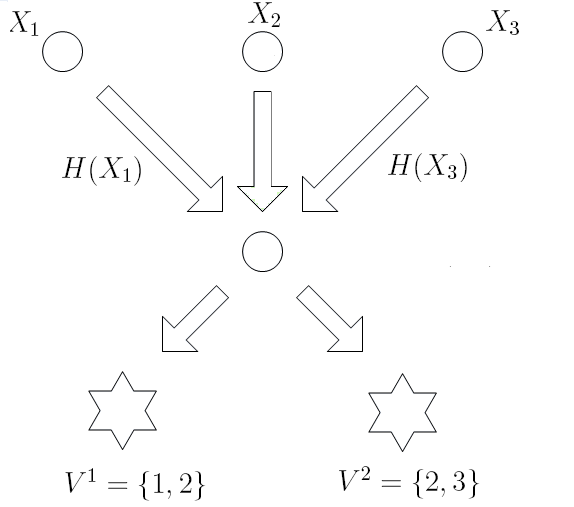}\caption{\label{fig:DIR_Motiv}Example to demonstrate the asymptotic gains
of DIR}

\end{figure}

Fig. \ref{fig:DIR_Motiv} depicts a simple network with 3 sources
$(X_{1},X_{2}$ and $X_{3})$ and two sinks $(S_{1}$ and $S_{2})$.
$S_{1}$ requests for $\{X_{1},X_{2}\}$ while $S_{2}$ requests for
$\{X_{2},X_{3}\}$. There is one intermediate node, c (called collector),
which serves the purpose of a simple router. The sources communicate
with the sinks only through the collector. This is a toy example for
a large sensor network with all intermediate nodes collapsed to a
single collector node. Note that we motivate our approach using the
loss-less setting (asymptotic behavior). But the practical design
(non-asymptotic) makes it feasible to work directly within the lossy
coding setting with loss-less as a special case. 

The collector has to transmit enough information to $S_{1}$ so that
it can decode both $X_{1}$ and $X_{2}$ and similarly enough information
to $S_{2}$ to decode $X_{2}$ and $X_{3}$. Hence the rates on the
edges $(c,S_{1})$ and $(c,S_{2})$ are at least $H(X_{1},X_{2})$
and $H(X_{2},X_{3})$, respectively. Let us say the weights on the
edges are such that, $w_{1,c}$,$w_{3,c}$$\ll$ $w_{2,c},w_{c,S_{1}}$,$w_{c,S_{2}}$.
This implies that $X_{1}$ and $X_{3}$ transmit data at rates $H(X_{1})$
and $H(X_{3})$, respectively. As source $X_{2}$ has to transmit
enough data for both the sinks to decode it losslessly: \begin{equation}
R_{2}\geq\max(H(X_{2}|X_{1}),H(X_{2}|X_{3}))\end{equation}
Conventional routing methods (designed for independent sources) do
not \textquotedbl{}split\textquotedbl{} a packet at an intermediate
node and hence would forward all the bits from $X_{2}$ to both the
sinks. This would mean sub-optimality on either one of the branches
$(c,S_{1})$ or $(c,S_{2})$ if $H(X_{2}|X_{1})\neq H(X_{2}|X_{3})$. 

But instead, let us now relax this restriction and allow the collector
to route only a subset of bits on each edge. Note that instead of
\textquotedbl{}splitting the packet\textquotedbl{} at the collector,
we could equivalently think of source $X_{2}$ transmitting 3 smaller
packets to the collector. First packet has rate $R_{2,\{1,2\}}$ bits
and is destined to both the sinks. Two other packets have rates $R_{2,1}$
and $R_{2,2}$ bits and are addressed to sinks $S_{1}$ and $S_{2}$,
respectively. Technically, in this case, the collector would just
have to route the received packets in a conventional manner. It can
be shown using random product binning arguments (refer to \cite{ITW10})
that the rate tuple $(R_{2,\{1,2\}},R_{2,1},R_{2,2})=(H(X_{2}|X_{1},X_{3}),I(X_{2},X_{3}|X_{1}),I(X_{2},X_{1}|X_{3}))$
is achievable and the rates on edges $(c,S1)$ and $(c,S2)$ achieve
their respective lower bounds.

We term such a routing mechanism, where each intermediate node in
a multi-hop network can \textit{route any subset} of the received
bits on each of the forward paths as \textquotedbl{}dispersive information
routing\textquotedbl{} (DIR). In the next section we propose a practical
integrated framework for efficient design of joint distributed source
coders and dispersive information routers that allows us to trade-off
the average reconstruction quality against the total communication
cost. Note the clear difference from network coding. DIR does not
require expensive coders at intermediate nodes, but rather can always
be realized using conventional routers with each source transmitting
multiple smaller packets into the network, intended to different subsets
of sinks. Also note that such a routing mechanism is inessential when
the sources are independent.

\subsection{Integrated DSC and Dispersive Routers\label{sub:Integrated-DSC-and}}

\begin{figure}
\centering\includegraphics[scale=0.32]{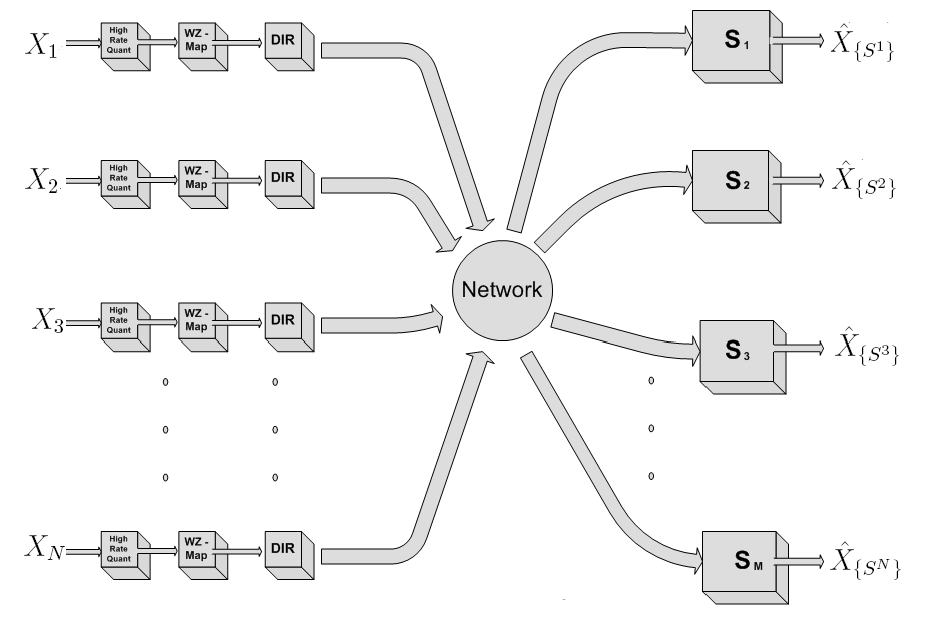}\caption{Integrated framework for DSC and DIR. The DIR modules decide the subset
of sinks each bit must be routed to. }

\end{figure}

Consider a network as formulated in subsection \ref{sub:DIR_Problem-Setup}
with each sink $S_{j}$ requesting for a subset $V^{j}\subseteq V$
of sources. Recall the encoding mechanism at each source. The random
variable $X_{i}$ is first quantized using the high rate quantizer
$\mathcal{H}_{i}$ and then mapped to the transmission index $I_{i}$,
using the WZ-map. Before we describe the proposed integrated framework
for DSC and DIR design, we first consider two standard approaches
for routing in a network and mention their drawbacks.

\subsubsection{Broadcasting}

The bits from all the sources are routed to all the sinks (broadcasted).
Then the decoder $\mathcal{D}_{ij}$ for source $X_{i}$ at sink $S_{j}$
can be expressed as:\begin{equation}
\mathcal{D}_{ij}:\mathcal{I}\rightarrow\hat{\mathcal{X}}_{ij}\subset\mathcal{R}\,\,\,\,\,\,\,\,\forall j,i\in V^{j}\end{equation}
Such a routing mechanism, though might provide good distortion performance,
is highly wasteful with respect to the communication cost.

\subsubsection{Conventional Routing}

All the bits transmitted by source $i$ are routed to its destination
sinks $S_{j}\in S^{j}$. If we use $\mathcal{I}_{S_{j}}=\prod_{i\in S^{j}}\mathcal{I}_{i}$,
to denote the set of all possible received symbols at sink $S_{j}$,
then the decoder $\mathcal{D}_{ij}$ can be expressed as:\begin{equation}
\mathcal{D}_{ij}:\mathcal{I}_{S_{j}}\rightarrow\hat{\mathcal{X}}_{ij}\subset\mathcal{R}\,\,\,\,\,\,\,\,\forall j,i\in V^{j}\end{equation}
As we saw in the previous sub-section, such a routing technique, wherein
fragmentation of the packets is not allowed, is (asymptotically) suboptimal
for communicating correlated sources over a network. Another major
drawback with such a routing technique is that, an unrequested but
correlated source may provide less expensive information about requested
sources. Hence, restricting each sink to receive packets only from
the requested sources could be highly suboptimal. 

Motivated by these drawbacks, the proposed dispersive information
routing paradigm allows for every sink to receive information from
all the sources, regardless of the request matrix. Moreover, we allow
for packet splitting at intermediate nodes, i.e., a sink may receive
only a subset of the bits transmitted by any given source. Essentially,
\textit{we allow each bit to be routed to any subset of the sinks}.
We introduce a new module at each encoder which decides the route
for each bit generated at that encoder. Note that if each bit follows
the route prescribed by the encoders, every intermediate node would
just be forwarding a subset of the received bits on each of the forward
paths. We call this module the {}``dispersive information router
module'' to indicate the routing mechanism it induces in the network.
We denote by $S=\{S_{1},S_{2}\ldots S_{M}\}$ the set of all sinks
and by $2^{S}$ the power set (set of all subsets) of $S$. Formally,
the router module at the $i^{th}$ encoder is given by:\begin{equation}
\mathcal{C}_{i}:\{1,\ldots,R_{i}\}\rightarrow2^{S}\end{equation}
and denote by $\mathcal{C}=\mathcal{C}_{1}\times\mathcal{C}_{2}\ldots\mathcal{C}_{N}$.
The router modules uniquely determine the set of all the received
bits at each sink. It is important to note the similarities between
the bit-subset selector module in Section \ref{fig:Proposed-setup-LSDQ}
and the dispersive information router modules here. In the former
case, the bit-subset selector decides the subset of received bits
to be used to estimate each source and the objective is to minimize
the average reconstruction distortion subject to a constraint on the
total decoder complexity. Whereas in the latter case, the dispersive
information router modules decide the subset of the transmitted bits
that are routed to each sink, to estimate the requested sources. The
objective is to design the router modules (jointly with the encoders
and the decoders) to minimize the average distortion, subject to a
constraint on the total communication cost. The underlying principles
in both these setting are exactly the same, though the explicit cost
functions are different. Hence, we use the same principles described
in Section \ref{sec:Algorithms-for-System} for the joint design of
encoders, routers and the reconstruction codebooks. Due to the close
relation between the two problems, we only state the Lagrangian cost
function to be optimized in the DIR framework and omit restating the
details of the design algorithms. 

The decoder at each sink is now modified to be the mapping:\begin{equation}
\mathcal{D}_{ij}:\mathcal{I}\times\mathcal{C}\rightarrow\hat{\mathcal{X}}_{ij}\subset\mathcal{R}\,\,\,\,\,\,\,\,\forall j,i\in V^{j}\end{equation}
The total communication cost $W$ of the system is given by:\begin{equation}
W=\sum_{i=1}^{N}\sum_{j=1}^{R_{i}}d_{i}^{*}\left(\mathcal{C}_{i}(j)\right)\label{eq:total_cost}\end{equation}
and the average reconstruction distortion is measured as:\begin{equation}
D=E\left[\sum_{j=1}^{M}\sum_{i=1}^{|V^{j}|}\gamma_{ij}d_{ij}(X_{i},\hat{X}_{ij})\right]\label{eq:}\end{equation}
where $d_{ij}:R\times R\rightarrow[0,1)$ are well-defined distortion
measures and $\gamma_{ij}$ are used to weigh the relative importance
of each source at each sink to the total distortion. If we specialize
the distortion metric to be the mean squared error and assume the
expectations to be approximated by empirical averages, the average
distortion is measured as,\begin{equation}
D=\frac{1}{|\mathcal{T}|}\left[\sum_{\mathbf{x}\in\mathcal{T}}\sum_{j=1}^{M}\sum_{i=1}^{|V^{j}|}\gamma_{ij}(x_{i}-\mathcal{D}_{ij}(I,\mathcal{C}))^{2}\right]\label{eq:distortion}\end{equation}
where $\mathbf{x}=\{x_{1}\ldots x_{N}\}$ and $I=[\mathcal{E}_{1}(x_{1}),\mathcal{E}_{2}(x_{2})\ldots\mathcal{E}_{N}(x_{N})]$
denotes the set of all bits being routed in the network. The trade-off
between the distortion and the communication cost is controlled using
a Lagrange parameter $\lambda\geq0$ to optimize the weighted sum
of the two quantities. From (\ref{eq:distortion}) and (\ref{eq:total_cost}),
the Lagrangian cost to be minimized is: \begin{eqnarray}
L=D+\lambda W\label{eq:Lagrangian}\end{eqnarray}
where $D$ and $W$ are given by (\ref{eq:distortion}) and (\ref{eq:total_cost}),
respectively. The objective is to find $\mathcal{E}_{i}'s$, $\mathcal{C}_{i}'s$
and $\mathcal{D}_{ij}'s$ that minimize $L$ for a given $\lambda$.
We again note that this optimization framework is exactly same as
that in Section \ref{sec:LSDQ_Proposed}, but with a different Lagrangian
cost and hence we omit restating the design algorithms. We also note
that, for large networks, it would be necessary to explicitly control
the decoder complexity while trading-off distortion and communication
cost. This could be easily performed by introducing multiple Lagrange
parameters to optimize the weighted sum of all the three quantities.
We omit the details here as it is a direct extension of the above
formulations. We note that the above approach for the design of the
dispersive information router modules is centralized in the sense
that the optimization is done offline, at a central location, before
the operation. In this paper, we aim to establish the potential gains
from using DIR in a practical setting, which in turn promotes future
research for developing efficient decentralized design strategies.

\subsection{Results for Dispersive Information Routing\label{sec:results}}

\selectlanguage{american}%
\begin{figure}
\selectlanguage{english}%
\centering\includegraphics[scale=0.27]{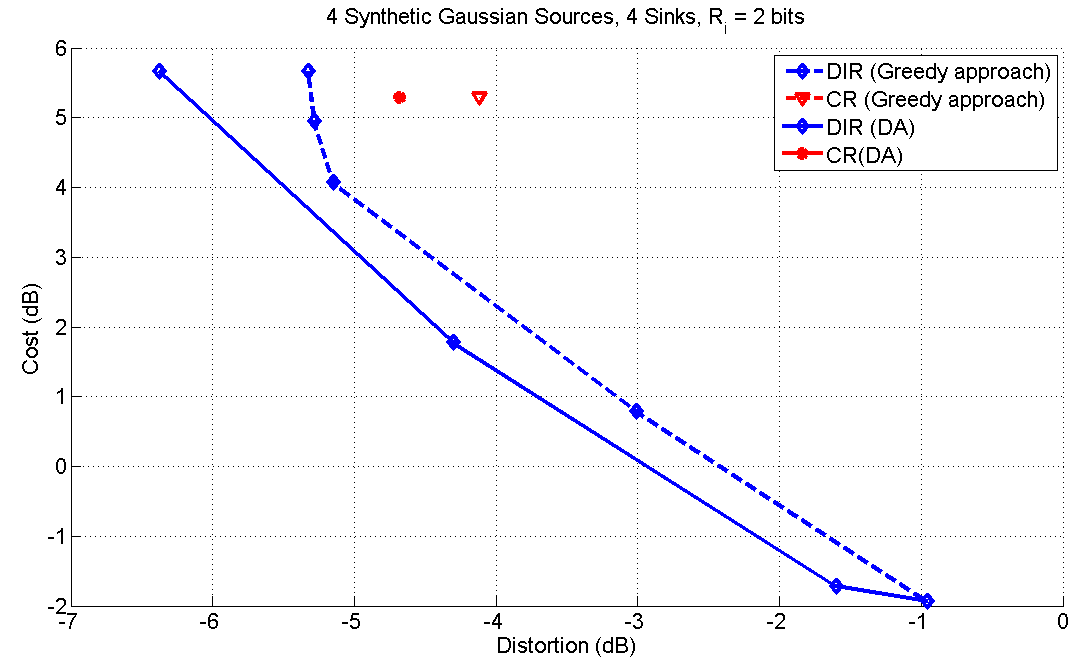}\foreignlanguage{american}{\caption{\selectlanguage{english}%
Cost versus distortion trade-off for 4 Randomly deployed Gaussian
sources and 4 sinks \label{fig:5_LSDQ-1}\selectlanguage{american}
}
}\selectlanguage{american}

\end{figure}

\begin{figure}
\selectlanguage{english}%
\centering\includegraphics[scale=0.27]{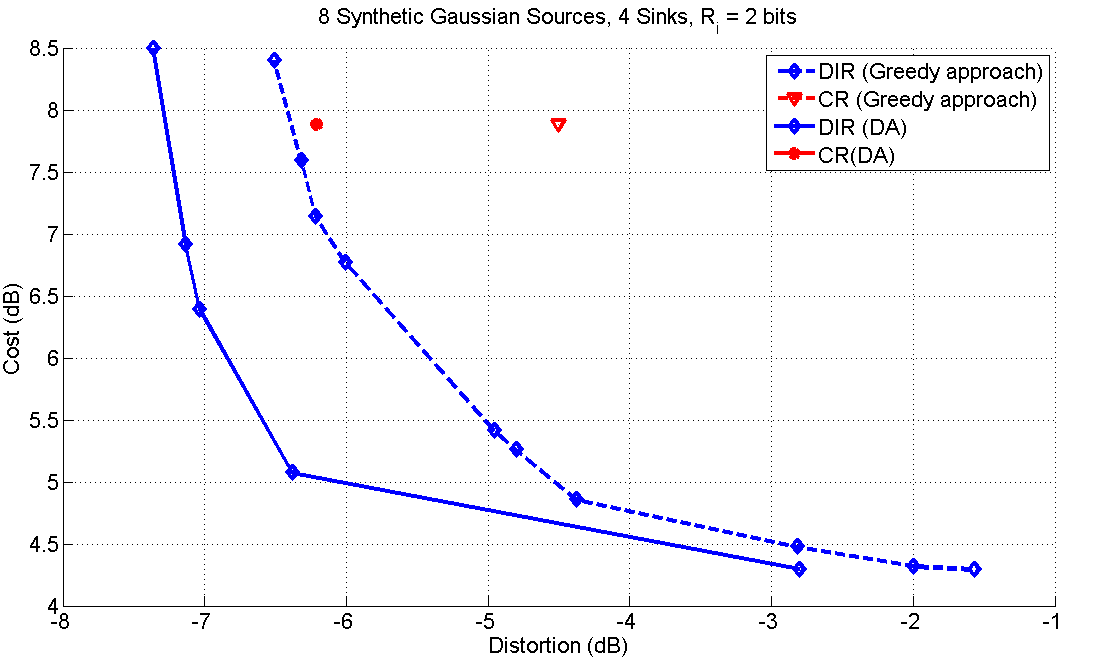}\foreignlanguage{american}{\caption{\selectlanguage{english}%
Cost versus distortion trade-off for 8 Randomly deployed Gaussian
sources and 4 sinks \foreignlanguage{american}{\label{fig:10_LSDQ-1}}\selectlanguage{american}
}
}\selectlanguage{american}

\end{figure}

\begin{figure}
\selectlanguage{english}%
\centering\includegraphics[scale=0.27]{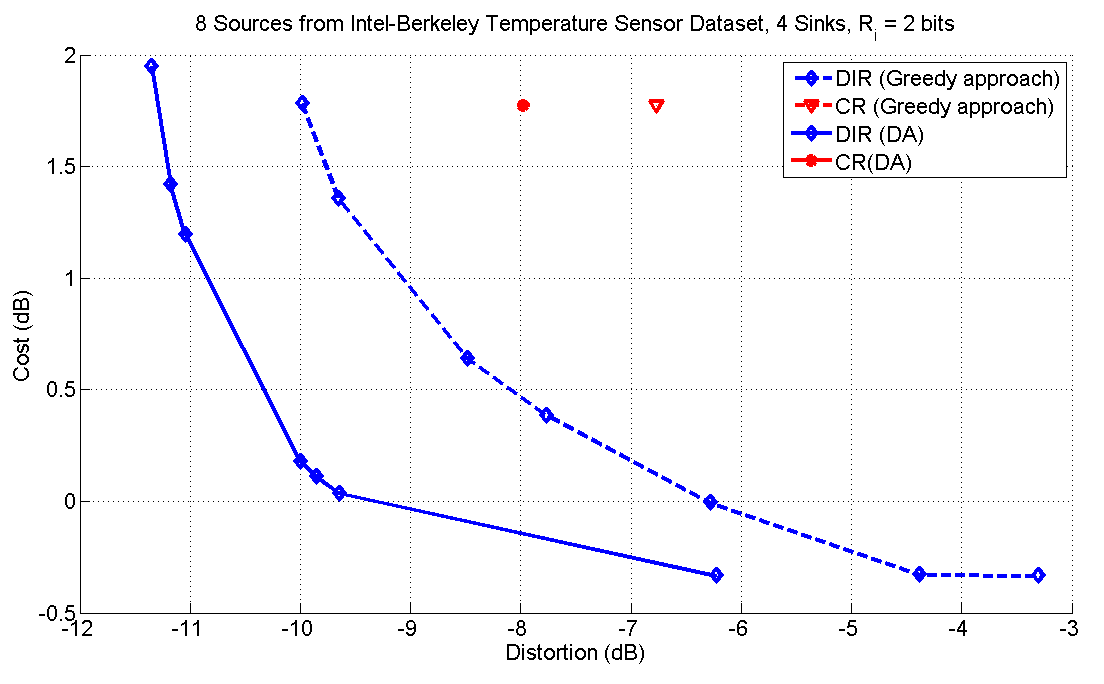}\foreignlanguage{american}{\caption{\selectlanguage{english}%
Cost versus distortion trade-off for\foreignlanguage{american}{ }Intel-Berkeley
temperature sensor dataset, 8 sources and 4 sinks\foreignlanguage{american}{\label{fig:Berk_LSDQ-1}}\selectlanguage{american}
}
}\selectlanguage{american}

\end{figure}

\selectlanguage{english}%
We considered three sensor network grids to demonstrate the gains
achievable during dispersive information routing compared to the conventional
routing techniques. We assumed 4 sinks for all three grids, placed
at the 4 corners of a square grid of length $d_{0}=100$. The first
two datasets were synthetic Gaussian datasets obtained by random deployment
of the sensors and intermediate nodes, to mimic a real world scenario
with inaccessible regions. The correlation between two sensors at
a distance $d$ was assumed to be $\rho^{\nicefrac{d}{d_{o}}}$. We
assumed $\rho=0.8$ for both these experiments. The first simulation
had 4 sources with 10 other intermediate nodes, while the second simulation
assumed 8 sources and 15 other intermediate nodes. For both these
simulations, the training set size was assumed to be $200,000$ samples.
The third dataset we considered consisted of readings from 8 temperature
sensors chosen randomly from the dataset collected by the Intel Berkeley
research center. All the other 46 sensors were regarded as intermediate
nodes with simple routing capabilities. The 4 sinks were assumed to
be at the four corners of the building. Again, half the samples were
used for training and the remaining half for testing. For a source
transmitting at a rate $R_{i}$, the high rate quantizer partitioned
the source space into $2^{R_{i}+3}$ regions, for e.g. if the source
rate was $2$ bits, $N_{i}=32$. For designs based on greedy-iterative
decent based methods, we report the best performance over several
random initializations (limited to 25). We used the square of the
distance between two sensors as the corresponding edge weight ($w_{e}$)
for all the simulations. The Steiner tree optimized costs $(d_{i}^{*})$
were computed before the design of the modules, using an optimal Steiner
tree algorithm. We note that the minimum Steiner tree problem is NP-hard
and requires approximate algorithms to solve in practice for larger
networks. 

An operational cost-distortion trade-off curve was obtained using
the proposed design techniques for all the three datasets. Figures
\ref{fig:5_LSDQ-1}, \ref{fig:10_LSDQ-1} and \ref{fig:Berk_LSDQ-1}
show the C-D curves for the three datasets, respectively. As a competitor,
we show that performance of optimally designed DSC for conventional
routing. As it is clearly evident, DIR gains close to $2$ dB in distortion
for the 4 source and $1.5$ dB in distortion for the 8 source synthetic
grids, respectively. For the real sensor network dataset, we see higher
gains of above $3$ dB compared to the conventional routing technique.
It is also evident that the DA based designs show significant reduction
in communication cost at a fixed reconstruction distortion. The gains
in communication cost due to DA based design for the three datasets
are $1$ dB, $2$ dB and $1.5$ dB, respectively.

\section{Conclusion}

We proposed a new decoding paradigm for large-scale distributed coding
which operates at practical codebook complexities. We formulated a
Lagrangian cost and proposed a design algorithm based on deterministic
annealing (DA) to optimize the performance trade-off between complexity
and distortion. We then pointed out close similarities between the
problem of large scale distributed coding and the problem of dispersive
information routing for communicating correlated sources across a
network. We used similar design principles proposed in the context
of large scale distributed coding for the design of an integrated
framework for distributed source coders and dispersive information
routers. Simulation results on both real and synthetic datasets show
considerable gains of the proposed approaches over conventional methods.

\bibliographystyle{unsrt}
\bibliography{Journal_Bibtex}

\end{document}